\documentclass{aa}
\usepackage[varg]{txfonts}

\usepackage{hyperref}

\usepackage{graphicx}
\usepackage{gensymb}
\usepackage{color}

\def\apjl{ApJL}

\title{Dynamical constraints on a dark matter spike at the Galactic Centre from stellar orbits}

\author{Thomas Lacroix}
\institute{Laboratoire Univers \& Particules de Montpellier (LUPM),
  CNRS \& Universit\'e de Montpellier (UMR-5299),
  Place Eug\`ene Bataillon,
  F-34095 Montpellier Cedex 05, France\\
\email{thomas.lacroix@umontpellier.fr}}

\begin{document}

\titlerunning{Dynamical constraints on a dark matter spike at the Galactic Centre from stellar orbits}
\authorrunning{T. Lacroix}

\abstract{In this work I use astrometric and spectroscopic data on the S2 star at the Galactic Centre (GC) up to 2016 to derive specific constraints on the size of a dark matter (DM) spike around the central supermassive black hole Sgr A*. These limits are the best direct constraints on a DM spike at the GC for non-annihilating dark matter and exclude a spike with radius greater than a few tens of parsecs for cuspy outer halos and a few hundred parsecs for cored outer halos.}

\keywords{Dark matter - Galactic Centre - celestial mechanics - black hole}

\maketitle

\section{Introduction}
\label{intro}

Dark matter profiles in the central regions of galaxies are poorly constrained at present and are the objects of intense debate. While observations seem to favour flat (cored) profiles, numerical simulations favour steeper profiles (cusps), leading to the cusp/core controversy (e.g. \citealp{deBlok2010} for a review). At subparsec scales, the dark matter (DM) distribution is even less constrained and can be significantly affected by the central supermassive black hole (SMBH). In particular, if the SMBH grows adiabatically, i.e.~on a much longer timescale than the dynamical timescale, the DM density is expected to be significantly enhanced (by up to 10 orders of magnitude at the very centre) in a region corresponding to the sphere of influence of the black hole (BH), typically at parsec scales for the Milky Way. This leads to a very sharp morphological feature referred to as a DM spike, corresponding to a DM profile going as $r^{-\gamma_{\mathrm{sp}}}$, with $\gamma_{\mathrm{sp}}$ typically between 2.25 and 2.5, depending on the slope of the initial DM halo \citep{spikeGS}. DM spikes are of particular interest in the context of indirect DM searches since they lead to very strong signatures of DM annihilation and allow us to probe weakly annihilating DM particles \citep{spikeGS,Regis&Ullio,Spike_GC_my_paper_no_a,M87_limits_my_paper,M87_EHT_my_paper,Fields2014,Shapiro2016}.

There is, however, considerable uncertainty on the formation and survival of DM spikes. In particular, the assumption of adiabaticity may not be verified in general. For instance, dynamical processes such as mergers can lead to weaker cusps \citep{Merritt2002}. However, binary scouring only occurs above parsec scales, while we are interested in the DM profile much closer in when studying the orbits of S stars, as discussed in the following. Moreover, the Milky Way (MW) is unlikely to have suffered such mergers in its recent past, as evidenced by the quiet history of the thick disk since the only major merger which occurred about 12 Gyr ago and is likely to have led to the formation of the bulge and the SMBH \citep{Wyse2001}. A weaker cusp is also formed if the BH does not grow exactly at the centre of the DM halo (within $\sim 50\ \rm pc$) \citep{Nakano1999,Ullio2001} or if the BH growth cannot be considered adiabatic \cite{Ullio2001}, but the actual impact of these effects on the MW is unclear. Moreover, dynamical heating in the central stellar core would also soften a spike \citep{Gnedin2004}. Another concern is that the non-observation of a stellar spike (recent results point to a softer stellar cusp than previously thought, with slope $\sim 1.15$; \citealt{Schoedel2017}) would rule out the existence of a DM spike. However, if the BH grows (for example by gas accretion) mostly before the nuclear star cluster forms in the spike region, then the DM and stellar profiles are decoupled. Additionally, the nuclear star cluster in the most accepted view is formed by merging globular clusters, as in \citet{Antonini2015}. This leads to different profiles for the DM and stellar distributions. Therefore, stars and DM essentially decouple, and the absence of a stellar spike in observations does not preclude the existence of a DM spike. On the other hand, additional dynamical processes can have the opposite effect of regenerating a spike, for example enhanced accretion of DM to counteract the depopulation of chaotic orbits in triaxial halos \citep{Merritt2004triaxial} or gravo-thermal collapse for self-interacting DM \citep{Ostriker2000}. 

As a result the unclear status of the inner DM profile of galaxies as discussed above calls for direct probes. In particular, there is still no definitive evidence either in favour of or against such a high concentration of DM either in the MW or in any other galaxy. This is due in particular to the small size of the regions involved. Probing such regions requires high angular resolution and astrometric precision to characterize the gravitational potential. However, the inner region of the MW offers a unique window on the DM distribution at the Galactic Centre (GC), thanks to the monitoring of the orbits of the  S stars within $\sim 1\, \rm arcsec$ of the central BH. In particular, since it is the closest star to the BH observed so far, the S2 star has been extensively studied through monitoring campaigns based on observations conducted with the Very Large Telescope (VLT) \citep{Schoedel2002,Gillessen2009a,Gillessen2009b,Gillessen2017,GRAVITYCollaboration2018} and the Keck observatory \citep{Ghez2005,Ghez2008,Boehle2016}.\footnote{Data from before 2002 were produced with the New Technology Telescope (NTT).} These series of observations have led to the reconstruction of the orbit of the star over roughly one and a half periods. In addition to tight constraints on the mass of the central SMBH, $M_{\mathrm{BH}}$, and its distance from Earth, $R_{0}$, these two groups have shown that only a small fraction (typically 1-2\%) of the mass of the SMBH can be in the form of an extended distribution \citep{Ghez2008,Gillessen2009a,Gillessen2009b,Boehle2016,Gillessen2017}. Other constraints have been obtained on an extended component by studying the corresponding reconstructed mass profile \citep{Hall2006} or the pericentre shift of S2 \citep{Zakharov2007,Iorio2013}.

Here I go a step further and I use astrometric and spectroscopic measurements of the orbit of S2 up to 2016 to set specific constraints on the DM distribution in the inner Galaxy. I present the first direct dynamical constraints from stellar orbits on the size of a DM spike, inside a DM halo constrained by larger scale kinematic data at kpc scales, for example from  maser observations. This is especially interesting for  non-annihilating or very weakly annihilating DM which is not expected to have significant observational signatures other than gravitational.

In Sect.~\ref{model} I describe the model along with the orbit-fitting procedure, before presenting my results in Sect.~\ref{results}. Finally, I conclude in Sect. \ref{conclusion}.

\section{Model and orbit-fitting procedure}
\label{model}

\subsection{Calibration: the point-mass case}

I rely on textbook results of standard mechanics in a central potential (e.g. \citealp{Bate1971}). I first recall the parameters of the problem in the BH-only case, which has an analytic solution, before moving on to the more general case of an extended mass distribution. The BH-only case serves as calibration for the orbit-fitting procedure.

The orbit-fitting procedure consists in reconstructing the time evolution of the position and velocity of the star on its orbit to determine the properties of the gravitational potential by fitting the parameters of the model to the data. In the case of one star orbiting a central point mass, the 13 parameters of the problem are the mass of the central object, here denoted  $M_{\mathrm{BH}}$, and its six phase-space coordinates, namely its distance $R_{0}$, its position on the sky ($\alpha_{\mathrm{BH}}$, $\delta_{\mathrm{BH}}$), and velocity ($v_{\alpha,\mathrm{BH}}$, $v_{\delta,\mathrm{BH}}$, $v_{\mathrm{r,BH}}$), as well as the six phase-space coordinates of the star. However, the orbit of the star is more readily characterized analytically in terms of the six standard orbital elements: the semi-major axis $a$ of the orbit, the eccentricity $e$, the time of pericentre passage $t_{\mathrm{P}}$, and three angles, namely the inclination $I$ of the orbital plane with respect to the plane of the sky, the longitude of the ascending node $\Omega$, and the angle $\omega$ between the directions of the ascending node and the pericentre.

Although the motion of Sgr A* with respect to the local standard of rest (LSR), defined as the circular velocity at the radius of the Sun, is expected to be very small \citep{Reid2004,Plewa2015}, its position ($\alpha_{\mathrm{BH}}$, $\delta_{\mathrm{BH}}$) on the plane of the sky at a reference time $t_{\mathrm{ref}}$ and its velocity ($v_{\alpha,\mathrm{BH}}$, $v_{\delta,\mathrm{BH}}$, $v_{\mathrm{r,BH}}$) relative to the LSR are unknown a priori and can be constrained through the orbit-fitting procedure. In practice,  the motion of the BH is accounted for through a linear term in the angular position of the star as a function of time. The reference time is taken to be 2009 yr \citep{Gillessen2017} for the data set up to 2016, and 2005.4 yr for the data set up to 2009 \citep{Gillessen2009b}.

In this work, I used the data from the NTT/VLT and Keck observatories compiled in \citet{Boehle2016,Gillessen2017}. In \citet{Gillessen2009b} the authors presented a robust method to consistently combine the two independent data sets for which the astrometric data feature a clear offset due to slight differences in the definition of the coordinate systems. More specifically, to account for the discrepancy between the two data sets, they introduced an offset in angular position ($\Delta \alpha$, $\Delta \delta$) and velocity ($\Delta v_{\alpha}$, $\Delta v_{\delta}$) on the plane of the sky to shift the Keck data back onto the VLT data. This was done by fitting the model with these 4 parameters in addition to the 13 parameters described before. In practice, this is achieved by shifting the observed right ascensions and declinations of the star measured with the Keck observatory by the quantities $\Delta \alpha + \Delta v_{\alpha}(t-t_{\mathrm{ref}})$ and $\Delta \delta + \Delta v_{\delta}(t-t_{\mathrm{ref}}),$ respectively. I repeated this procedure here for the combined data set.

Throughout this work, I derived the posterior probability density function of model parameters using \textsf{PyMultiNest} \citep{Buchner2014}, which relies on the \textsf{MultiNest} code \citep{Feroz2009} based on the multimodal nested sampling Monte Carlo technique \citep{Feroz2008}. Multimodal nested sampling is particularly suitable for studying high-dimensional parameter spaces with possible degeneracies between parameters. The likelihood combines the data on right ascension, declination, and radial velocity of S2. I used uniform priors for all parameters except the position and velocity of the BH, for which I took Gaussian priors based on the results from \citet{Plewa2015}:  $(\alpha_{\rm BH},\delta_{\mathrm{BH}}) = (0,0) \pm (0.2,0.2)\, \rm mas$ at $t_{\mathrm{ref}} = 2009\, \rm yr$ and $(v_{\alpha,\mathrm{BH}},v_{\delta,\mathrm{BH}}) = (0,0) \pm (0.1,0.1)\, \rm mas\, yr^{-1}$. I recovered the best-fit parameters and errors from \citet{Gillessen2017}, as illustrated by the marginalized posterior distributions, for the BH-only case and the full VLT data set up to 2016 (see Fig.~\ref{corner_plot_BH_only_VLT_2016} in Appendix \ref{best_fit}). I also recovered the best-fit model for the combined VLT+Keck data set, using the prescriptions of \citet{Gillessen2009b} for the priors on $\Delta \alpha$, $\Delta \delta$, $\Delta v_{\alpha}$, and $\Delta v_{\delta}$. This served as a consistency check of the analysis chain, which was then applied to the study of the impact of a DM spike on the orbit of S2.

\subsection{Extended mass}

\begin{figure*}[t!]
\centering
\includegraphics[width=0.495\linewidth]{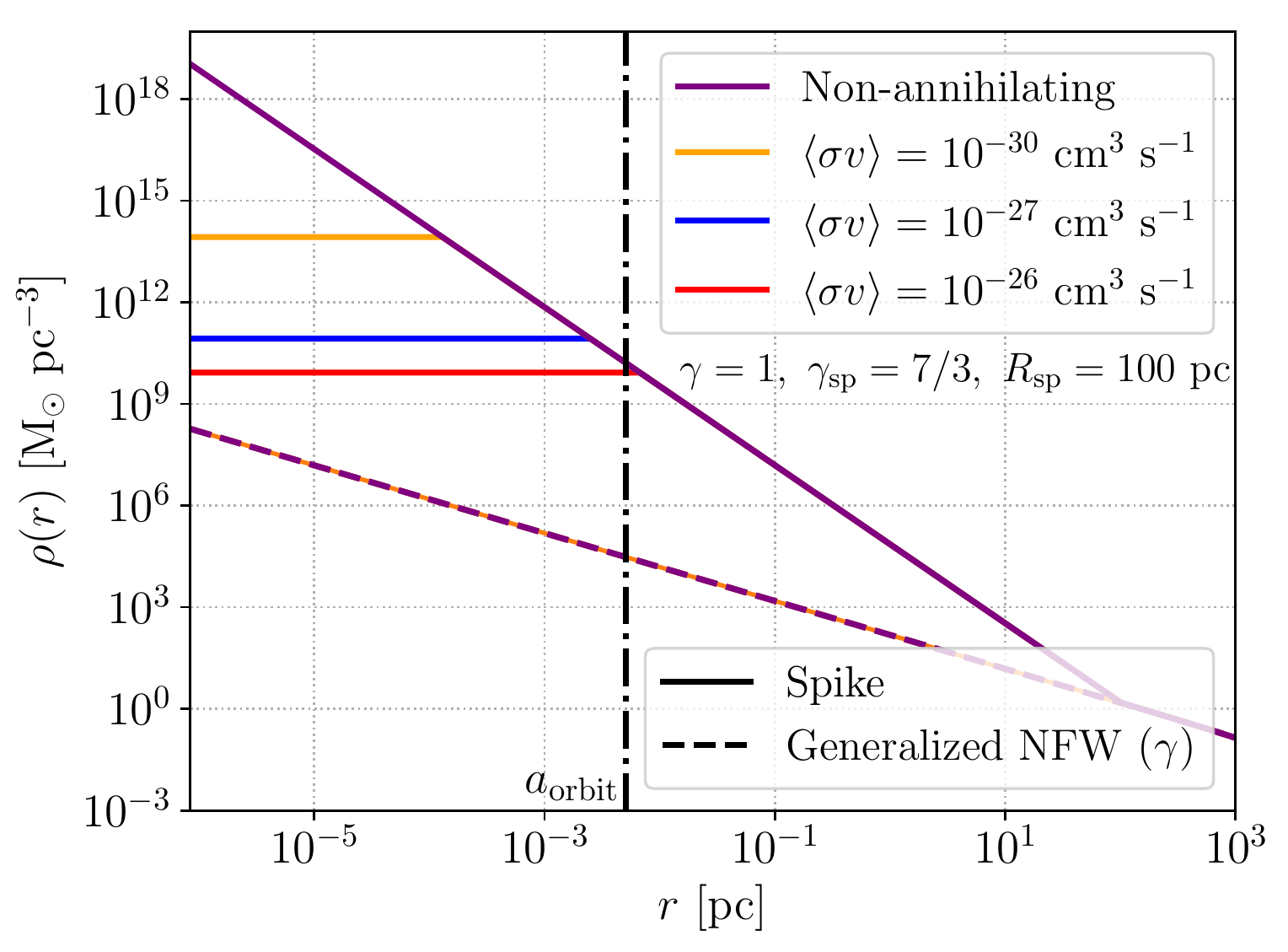} \hfill \includegraphics[width=0.495\linewidth]{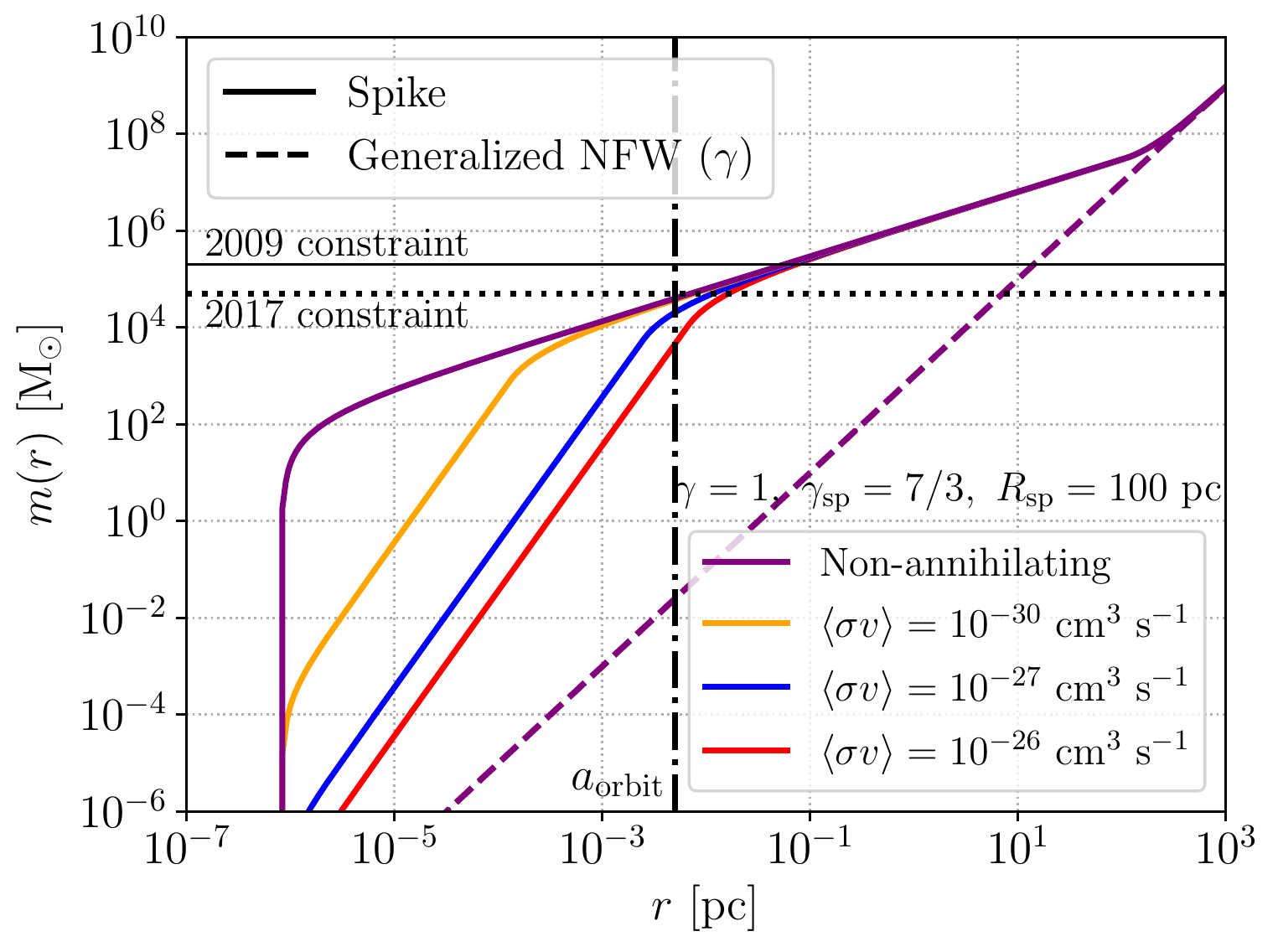} 
\caption{\label{density_profiles}Left panel: Density profiles for a generalized NFW halo (dashed lines) with $\gamma = 1$ and the same halo with a spike in the central region (solid lines) for a non-annihilating DM candidate (purple) and a self-annihilating 1 TeV DM candidate with $\left\langle \sigma v \right\rangle \sim 10^{-30}\, \rm cm^{3}\, s^{-1}$, $\left\langle \sigma v \right\rangle \sim 10^{-27}\, \rm cm^{3}\, s^{-1}$, and $\left\langle \sigma v \right\rangle \sim 10^{-26}\, \rm cm^{3}\, s^{-1}$ (orange, blue, and red, respectively). Right panel: Corresponding mass profiles, with the same line styles as in the left panel. The profiles are shown for illustration purposes for a spike radius $R_{\mathrm{sp}} \sim 100\, \rm pc$  corresponding to the 99.7\% upper limit  from deviations of the BH-only orbit using the VLT data (see Sect.~\ref{results}). The horizontal solid and dotted lines represent the combined 2009 and 2017 constraints, respectively. The vertical dot-dashed line marks the characteristic size of the orbit of S2.}
\end{figure*}

For the extended DM mass distribution, I consider two scenarios: the general case of a non-annihilating cold dark matter (CDM) candidate, and the more specific case of self-annihilating DM, applicable to candidates like weakly interacting massive particles (WIMPs). For non-annihilating DM the spike goes way inside the orbit of S2, down to the close vicinity of the SMBH \citep{Sadeghian2013}:
\begin{align}
\rho(r) =
\begin{cases}
0 & r < 2R_{\mathrm{S}} \nonumber \\
\rho_{\mathrm{halo}}(R_{\mathrm{sp}}) \left( \dfrac{r}{R_{\mathrm{sp}}} \right)^{-\gamma_{\mathrm{sp}}} & 2R_{\mathrm{S}} \leqslant r < R_{\mathrm{sp}} \nonumber \\
\rho_{\mathrm{halo}}(r) & r \geqslant R_{\mathrm{sp}},
\end{cases}
\end{align}
where $R_{\mathrm{sp}}$ is the radial extension of the spike, and the halo profile is assumed to be given by a generalized Navarro-Frenk-White (NFW) profile characterized by a slope index $\gamma$,
\begin{equation}
\rho_{\mathrm{halo}}(r) = \rho_{\mathrm{s}} \left( \dfrac{r}{r_{\mathrm{s}}} \right)^{-\gamma} \left( 1 + \dfrac{r}{r_{\mathrm{s}}} \right)^{\gamma-3},
\end{equation}
where $r_{\mathrm{s}}$ is the scale radius, and the scale density $\rho_{\mathrm{s}}$ is related to the local density $\rho_{\odot}$ via
\begin{equation}
\rho_{\mathrm{s}} = \rho_{\odot} \left( \dfrac{R_{0}}{r_{\mathrm{s}}} \right)^{\gamma} \left( 1 + \dfrac{R_{0}}{r_{\mathrm{s}}} \right)^{3-\gamma}.
\end{equation}

More specifically, the idea is to consider the various DM halos corresponding to the dynamically constrained Milky Way mass models from the analysis of \citet{McMillan2017}, and determine the maximum size of a DM spike inside that halo that does not cause a significant departure from the best-fitting BH-only orbit. The associated values of the local density, scale radius, and $R_{0}$ from the analysis of \citet{McMillan2017} are summarized in Table \ref{halo_params}, in Appendix \ref{mass_models}.\footnote{The best-fit values of $R_{0}$ from \citet{McMillan2017} are consistent with the values obtained with the orbit-fitting procedure. I do not keep $R_{0}$ in the generalized NFW profile as a free parameter, but  use it to fix the normalization of the halo profile in a way that is consistent with \citet{McMillan2017}. $R_{0}$ is only kept free in the position and velocity of the star.}

For self-annihilating DM, the inner region of the DM spike is depleted since the DM density is so high that DM particles annihilate more efficiently. This results in a plateau of density $\rho_{\mathrm{sat}} = m_{\mathrm{DM}}/(\left\langle \sigma v \right\rangle t_{\mathrm{BH}})$,
where $m_{\mathrm{DM}}$ is the mass of the DM candidate, $\left\langle \sigma v \right\rangle$ is the velocity-averaged annihilation cross section, and $t_{\mathrm{BH}}$ is the age of the central SMBH, which I take conservatively to be $\sim 10^{10}\, \rm yr$. The saturation plateau extends to a radius
\begin{equation}
\label{R_sat}
R_{\mathrm{sat}} = R_{\mathrm{sp}} \left[ \dfrac{\rho_{\mathrm{s}}}{\rho_{\mathrm{sat}}} \left( \dfrac{R_{\mathrm{sp}}}{r_{\mathrm{s}}} \right)^{-\gamma}  \right]^{1/\gamma_{\mathrm{sp}}}. 
\end{equation}


The cases of non-annihilating and self-annihilating DM are both illustrated in Fig.~\ref{density_profiles}. Shown are the density profiles for regular NFW-like halos and halos with a spike in the central region for non-annihilating DM and a self-annihilating 1 TeV DM candidate with three values of $\left\langle \sigma v \right\rangle$ (see figure for details). The corresponding mass profiles are shown in the right panel of Fig.~\ref{density_profiles} with the same line styles. The profiles are illustrated with $\gamma = 1$, which gives a spike slope $\gamma_{\mathrm{sp}} = 7/3$, and a spike radius $R_{\mathrm{sp}} \sim 100\, \rm pc$ which corresponds to the 99.7 \% upper limit I obtain from deviations of the BH-only orbit, as discussed in Sec.~\ref{results}. For the self-annihilating case, the values of the cross section are chosen to illustrate the point at which the annihilation plateau becomes as big as the characteristic size of the orbit, given by the semi-major axis constrained to be of the order of 5 mpc by the orbit-fitting procedure. For $m_{\mathrm{DM}} \sim 1\, \rm TeV$ and $\left\langle \sigma v \right\rangle \sim 10^{-26}\, \rm cm^{3}\, s^{-1}$, the mass enclosed inside the orbit is significantly reduced with respect to the case of non-annihilating or very weakly annihilating DM, down to values much smaller than a few percent of the BH mass, making deviations from the BH-only orbit undetectable.\footnote{These considerations can be extended to other values of $m_{\mathrm{DM}}$ via Eq.~(\ref{R_sat}).} 

In the absence of a spike, the DM halo has a negligible impact on the orbit of S2, as illustrated in the right panel of Fig.~\ref{density_profiles}, due to a much smaller mass enclosed in the orbit. Moreover, for completeness I have also considered the effect of a realistic stellar profile $\rho_{\mathrm{star}} \propto r^{-\gamma_{\mathrm{star}}}$, with $\gamma_{\mathrm{star}} \sim 1.15$ and $\rho_{\mathrm{star}}(1\, \rm pc) = 1.5 \times 10^{5}\, \rm M_{\odot}\, pc^{-3}$ \citep{Schoedel2017}. However, the corresponding mass enclosed in the orbit is about three orders of magnitude below the critical mass needed to have an impact on the orbit. The same conclusion applies to the stellar bulge profile from \citet{McMillan2017}.

\begin{figure*}[t!]
\centering
\includegraphics[width=0.33\linewidth]{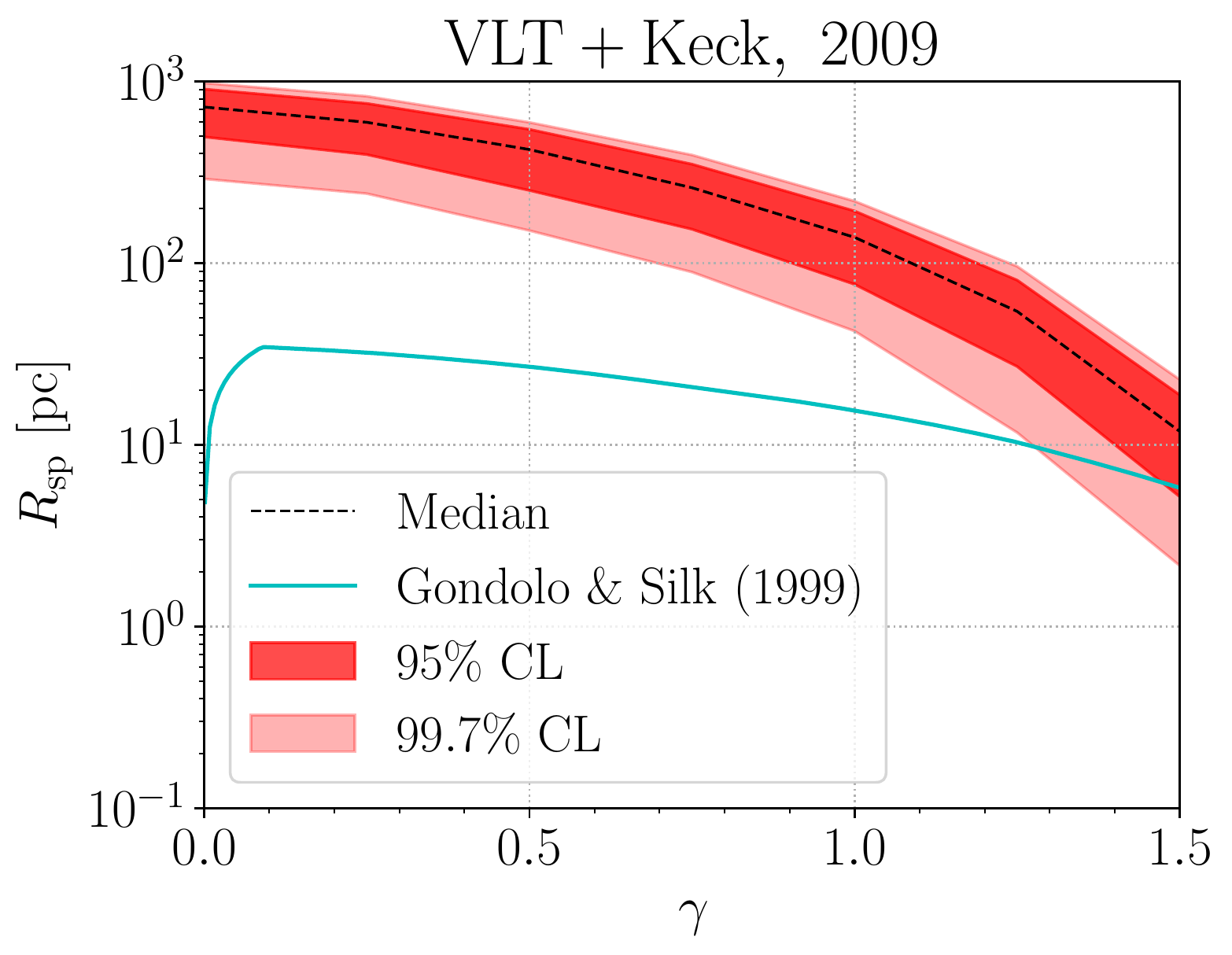} \includegraphics[width=0.33\linewidth]{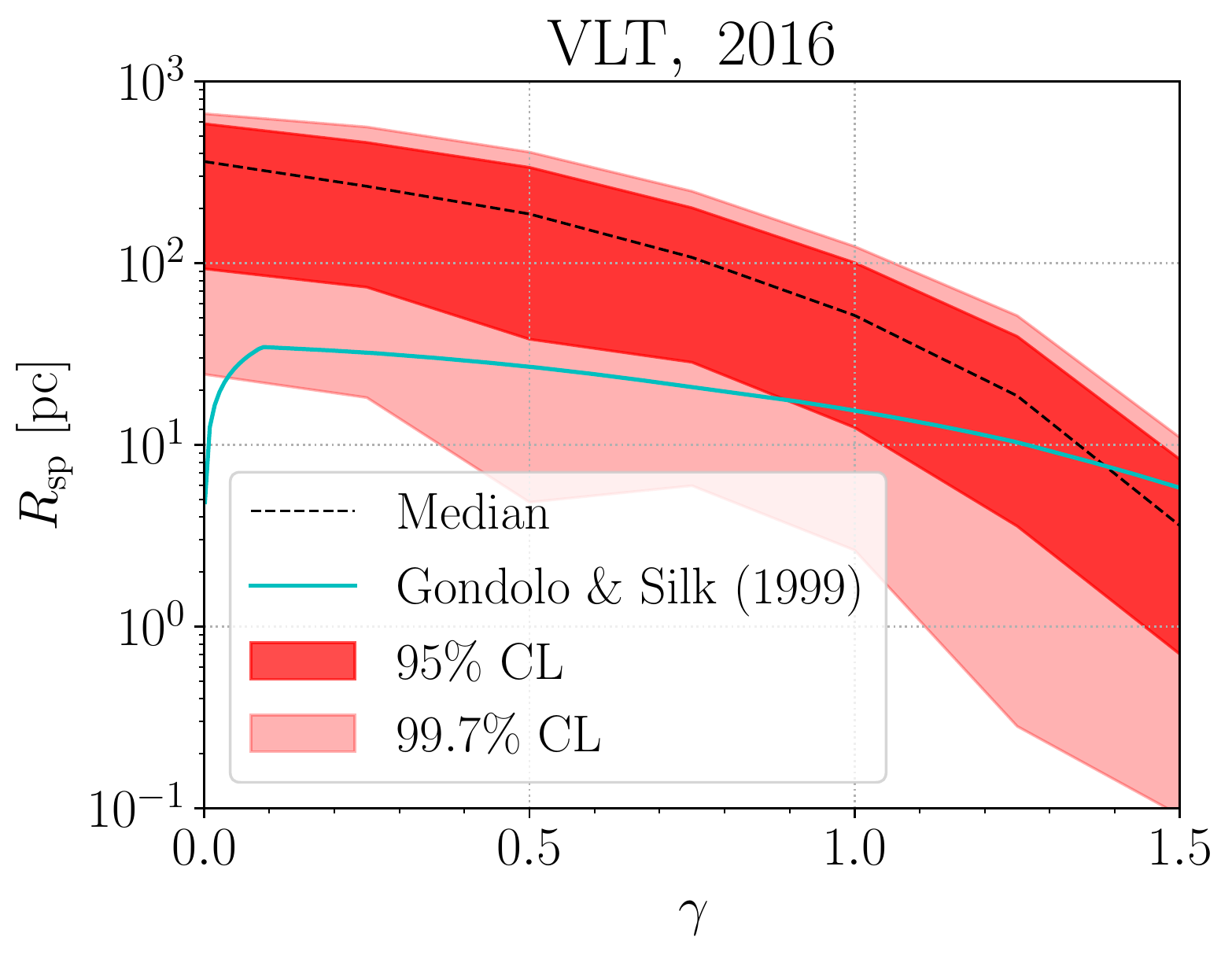} \includegraphics[width=0.33\linewidth]{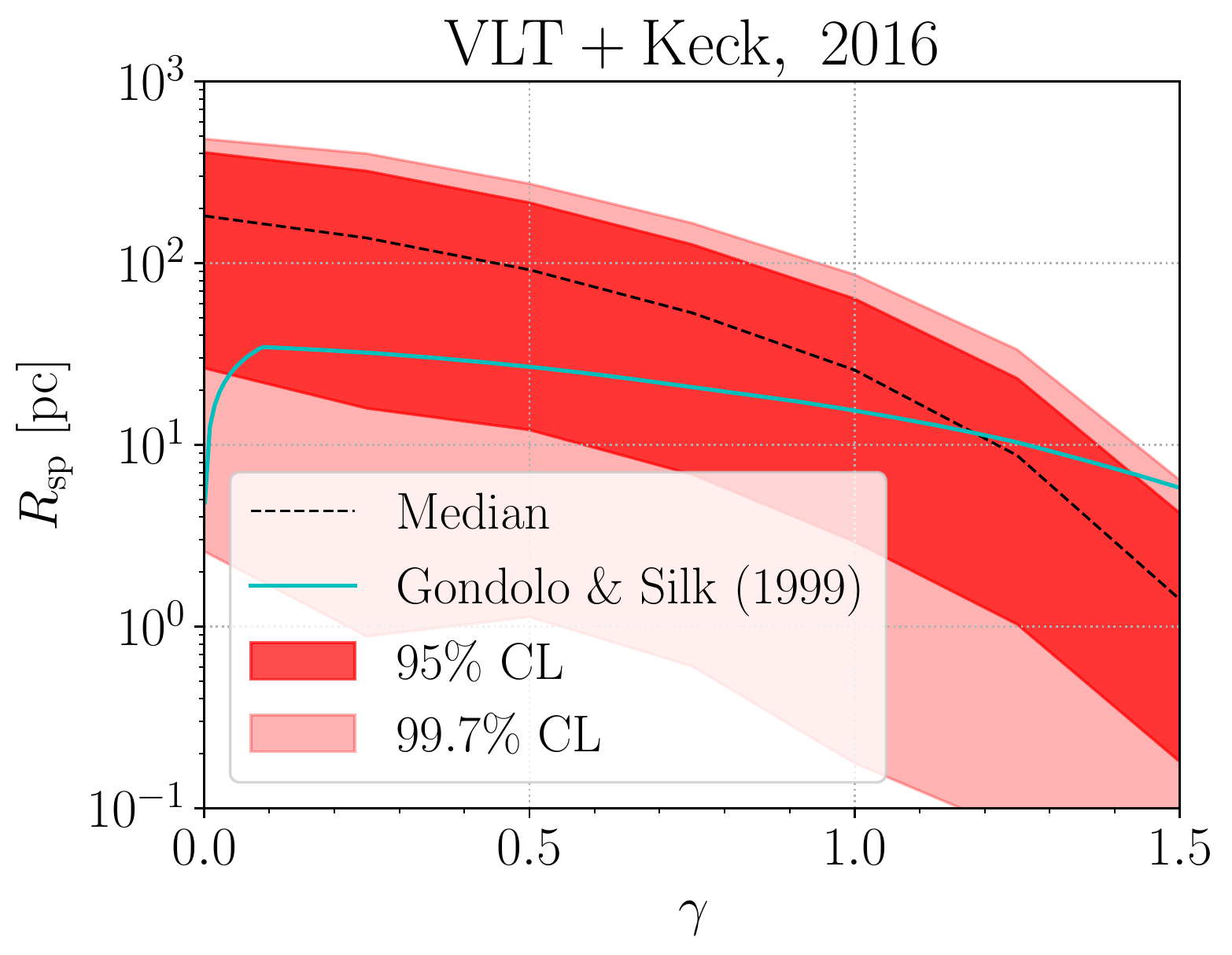} 
\caption{\label{upper_limits}Median of the marginalized posterior distribution of the spike radius $R_{\mathrm{sp}}$ (black dashed), and associated 95\% (red shaded) and 99.7\% (light red shaded) confidence contours as a function of the slope $\gamma$ of the corresponding dynamically constrained outer halos, using the combined VLT+Keck data up to 2009 (left panel), VLT-only data up to 2016 (middle panel), and VLT+Keck data up to 2016 (right panel). The upper right white regions correspond to excluded values of the spike radius. The lower left white regions are outside the 99.7\% best-fit contours, but cannot robustly be interpreted in terms of exclusion of the associated values of the spike radius since the Bayesian evidence does not favour the presence of a spike over the BH-only model. The contours are derived for values of the slope of the outer halo dynamically constrained by \citet{McMillan2017}, and linearly interpolated to get smooth curves. The prediction from \citet{spikeGS} is also shown as a benchmark model (cyan solid).}
\end{figure*}

In the general case of an extended mass distribution around the central point mass, the orbit model is no longer analytic and one must rely on numerical tools to solve the equations of motion. First the polar radius $r(t)$ is determined with Newton's second law in the Galilean frame of the BH, 
\begin{equation}
\label{pfd}
\ddot{r} - \dfrac{L^{2}}{r^{3}} = - \dfrac{G M_{\mathrm{BH}}}{r^{2}} - \dfrac{\mathrm{d}\Phi_{\mathrm{ext}}}{\mathrm{d}r} - \dfrac{\mathrm{d}\Phi_{\mathrm{S}}}{\mathrm{d}r},
\end{equation}
where $L \equiv r^{2}\dot{\theta}$ is the angular momentum modulus, $G$ is the gravitational constant, $\Phi_{\mathrm{ext}}$ is the potential created by the extended mass, and $\Phi_{\mathrm{S}}$ accounts for the effect of Schwarzschild precession induced by the BH.\footnote{Although the current data on S2 are not yet sensitive to relativistic effects \citep{Gillessen2017}, I  include this post-Newtonian precession effect for completeness since it is partly degenerate with the precession caused by the extended mass and as such can mildly affect the limits  set on the DM profile. Other relativistic effects such as gravitational redshift essentially affect radial velocities, which are much less tightly constrained by observations than the position of the star, as discussed in \citet{Gillessen2009a}.}
The initial conditions $r_{0} \equiv r(t_{0})$ and $\dot{r}_{0} \equiv \dot{r}(t_{0})$ need to be specified, where $t_{0}$ is chosen as the first epoch in the data, namely $t_{0} = 1992.224\ \rm yr$. For given values of $r_{0}$ and $\dot{r}_{0}$, I use the \textsf{odeint} Python routine to solve for $r(t)$. Once $r(t)$ is known, $\theta(t)$ is obtained via
\begin{equation}
\theta(t) = \theta_{0} + \int_{t_{0}}^{t} \! \dfrac{L}{r(t')^{2}} \, \mathrm{d}t',
\end{equation}
where $\theta_{0} \equiv \theta(t_{0})$ and $L = r_{0}^{2}\dot{\theta}_{0}$, with $\dot{\theta}_{0} \equiv \dot{\theta}(t_{0})$. Orbital elements no longer characterize the orbit but only an osculating orbit in the general case of an extended mass, so they cannot be used to parametrize the problem. The free parameters for the star are now the initial conditions $r_{0}$, $\theta_{0}$, $\dot{r}_{0}$, $\dot{\theta}_{0}$, as well as $I$ and $\Omega$, which still characterize the plane of the orbit of the star. The parameters of the BH do not change.\footnote{In order to keep the same coordinate system as in the BH-only case, I fixed the value of $\omega$, which is no longer a free parameter since the pericentre is not defined in general, to the best-fit value in the BH-only case.}

\section{Results}
\label{results}

In this section, I present constraints on the size of a DM spike as a function of the slope of the DM halo obtained with the multimodal nested sampling analysis implemented in \textsf{PyMultinest}. It should be noted that the nested sampling procedure does find a non-zero best-fit value for the spike radius $R_{\mathrm{sp}}$ for all values of $\gamma$. However, the very mild increase in Bayesian evidence when adding a DM spike, which remains smaller than $\Delta \ln Z \approx 3$, is insufficient to claim any preference for the BH+spike model \citep{Bayes_factors}. As a result, the data are consistent with the BH-only model. Nevertheless, it is still possible to exclude large values of the spike radius that would lead to a large DM mass inside the orbit, and thus to large deviations of the orbit of S2.

The resulting 95\% and 99.7\% confidence contours in the $\gamma$-$R_{\mathrm{sp}}$ plane are shown in Fig.~\ref{upper_limits} for the case of non-annihilating DM, for the dynamically constrained halo profiles from \citet{McMillan2017}, and using the combined VLT+Keck data sets up to 2009 (left panel), the VLT-only 2016 data set up to 2016 (middle panel), and the combined VLT+Keck data up to 2016 (right panel). The contours are computed for the values of halo slope $\gamma$ for which \citet{McMillan2017} derived constraints from various data sets including maser observations, namely $\gamma = 0,\, 0.25,\, 0.5,\, 0.75,\, 1,\, 1.25,\, 1.5$. I interpolated the results to obtain a smooth limit. \footnote{Accounting for the uncertainty on the local DM density $\rho_{\odot}$ from \citet{McMillan2017} only leads to a 4\% variation in the limits on $R_{\mathrm{sp}}$.} Using the combined VLT+Keck data set up to 2016, I exclude at the 99.7\% confidence level a DM spike with a spatial extension larger than 90 pc for an outer halo with $\gamma = 1$, and larger than 6 pc for an outer halo with $\gamma = 1.5$. For the combined 2009 and the VLT-only 2016 data sets, the limits are about a factor of 2 weaker. The analysis of \citet{McMillan2017} seems to favour cuspy halos ($\gamma \sim 1$), whereas a recent study on the dynamics of the Galactic bar favours a DM halo with slope $\gamma < 0.6$ \citep{Portail2017}. For such cored halos, our constraints are weaker, with a maximum spike radius of a few hundred pc. Nevertheless, these limits are the first direct constraints on a DM spike at the GC, valid for non-annihilating DM. This is especially interesting because it is applicable to any CDM candidate with no significant annihilation cross section. 

It should be noted that the limit  already improves by about a factor of 2 when going from the combined 2009 data set to the combined 2016 one. Significant additional improvements can thus be expected on the constraints with the more recent S2 data.

Constraints on an extended mass component can also be expressed more generally in terms of the total extended mass $M_{\mathrm{ext}}$ inside the characteristic size of the orbit. At a 99.7\% confidence level, for the combined data up to 2009, I find $M_{\mathrm{ext}} \lesssim 2 \times 10^{5}\, \rm M_{\odot}$, corresponding to $\sim 4\% M_{\mathrm{BH}}$, while for the complete data set up to 2016, I obtain $M_{\mathrm{ext}} \lesssim 4$--$5 \times 10^{4}\, \rm M_{\odot}$, corresponding to $\sim 1\% M_{\mathrm{BH}}$. These results are consistent with the upper limits on a general extended mass component from \citet{Gillessen2009b,Gillessen2017}. These limits are illustrated in the right panel of Fig.~\ref{density_profiles} by the horizontal dotted and solid black lines.

For self-annihilating DM, the constraints derived here are valid for $\left\langle \sigma v \right\rangle < 10^{-30}\, \rm cm^{3}\, s^{-1}$ for a benchmark particle mass $m_{\mathrm{DM}} = 1\, \rm TeV$. As shown in the right panel of Fig.~\ref{density_profiles}, for the same candidate mass, the mass enclosed in a sphere of radius of  the characteristic size of the orbit of S2---typically 5 mpc---is decreased by about a factor of 2 for $\left\langle \sigma v \right\rangle \sim 10^{-27}\, \rm cm^{3}\, s^{-1}$ and by about a factor of 20 for $\left\langle \sigma v \right\rangle \sim 10^{-26}\, \rm cm^{3}\, s^{-1}$ with respect to the mass in the absence of annihilation. As a result, the upper limits on $R_{\mathrm{sp}}$ are weakened by about a factor of 10 for $\left\langle \sigma v \right\rangle \sim 10^{-27}\, \rm cm^{3}\, s^{-1}$, while no constraints can be set for $\left\langle \sigma v \right\rangle \sim 10^{-26}\, \rm cm^{3}\, s^{-1}$. More generally, when the radius $R_{\mathrm{sat}}$ of the saturation plateau due to self-annihilations (see Eq.~(\ref{R_sat})) reaches the size of the orbit of S2, the DM mass enclosed inside the orbit becomes too small to induce significant deviations from the BH-only orbit.

For illustration, the prescription from \citet{spikeGS} is also shown in Fig.~\ref{upper_limits} (cyan solid line). This corresponds to a spike radius defined by
\begin{equation}
R_{\mathrm{sp}}^{\mathrm{GS}} = \alpha_{\gamma} r_0 \left( \frac{M_{\mathrm{BH}}}{\rho_{0}r_0^{3}} \right) ^{\frac{1}{3-\gamma}},
\end{equation}
with $\alpha_{\gamma} \approx 0.293\gamma^{4/9}$, which only differs from $\sim 0.1$ for $\gamma \ll 1$. It should be noted that these predictions are indicative and can be significantly affected by the various dynamical processes discussed in Sec.~\ref{intro}. For very cuspy halos ($\gamma \approx 1.5$), the combined 2016 data already exclude the prediction from \citet{spikeGS} at a 95\% confidence level.

\section{Conclusion and outlook}
\label{conclusion}

In this work, I have used an orbit-fitting procedure similar to those developed in \citet{Gillessen2017,Boehle2016} to derive specific constraints on the size of a DM spike for given outer DM halos dynamically constrained by larger scale observations. These limits are the best direct constraints on a DM spike at the GC for non-annihilating DM and exclude a spike with radius greater than a few tens of pc for cuspy outer halos and a few hundred pc for cored outer halos.

The addition of the 2017--2018 data from the VLT, which has monitored the pericentre passage of S2 \citep{GRAVITYCollaboration2018}, will make these constraints significantly more stringent, especially thanks to the impressive capabilities of the imaging NACO instrument, the SINFONI spectrometer, and the exquisite astrometric precision of the GRAVITY instrument. However, I postpone the study of the subsequent constraints on the DM profile at the GC to a future work since  additional subtleties related to the relativistic effects that have been detected using the new data \citep{GRAVITYCollaboration2018} may appear, and this warrants a dedicated study. The problem is also complicated further by having to model two pericentre passages when accounting for the entire data set since 1992, which increases the computing time.

Additional improvements on the data could lead to  even stronger constraints on the very inner DM profile at the GC. Firstly, in principle, S stars located further out than S2 would be more suited to probe the extended DM distribution for which the mass increases with radius. However, this comes at the price of longer periods, so that unlike S2, no additional stars have been monitored for about 1.5 periods. As a result, our constraints do not improve when including other stars further out such as S1 or S13 for which no significant precession is detectable yet. However, the situation will change when complete orbits are recorded for these stars. In addition,  even more accurate astrometric and spectroscopic data will be instrumental to further improve upon these constraints. In particular, a 30 m extremely large telescope (ELT) should be able to probe an extended mass component as low as a few $10^{3}\, M_{\odot}$, i.e. about one order of magnitude better than the current sensitivity, after only 10 years of observation \citep{Weinberg2005}. Moreover, an ELT would be able to break the degeneracy between relativistic effects and precession from an extended mass component. This would translate into sensitivity to DM spikes as small as a few pc even for cored outer halos, and even smaller for steeper halos.

\begin{acknowledgements}

I thank Benjamin Jaillant, Julien Lavalle, Vivian Poulin, and Martin Stref for fruitful discussions on this topic, and the anonymous referee for the very useful comments and suggestions. My work is supported by CNRS-IN2P3. I also acknowledge support from the European Union’s Horizon 2020 research and innovation program under the Marie Sk\l{}odowska-Curie grant agreements No 690575 and No 674896, in addition to recurrent institutional funding by CNRS-IN2P3 and the University of Montpellier.

\end{acknowledgements}

\bibliographystyle{aa} 
\bibliography{/Users/thomaslacroix/Documents/bibliography/biblio_bibtex/biblio}

\appendix

\section{Constrained DM halo models used in this work}
\label{mass_models}

\begin{table}[h!]
\caption{\label{halo_params}Parameters of the generalized NFW profiles for the Milky Way mass models constrained by the analysis of  \citet{McMillan2017}. The values of $R_{0}$ given in this table are only used to fix the normalization of the outer halo profile and do not exactly coincide with the radial coordinate of the BH at the GC.}
\centering
\begin{tabular}{c|c|c|c}
\hline \hline
 $\gamma$ & $\rho_{\odot}\ \rm [M_{\odot}\, pc^{-3}]$ & $R_{0}\ \rm [kpc]$ & $r_{\mathrm{s}}\ \rm [kpc]$\\
 \hline
0 & $ 0.0103 \pm 0.0009 $ & 8.21 & 7.7 \\
\hline
0.25 & $0.0100 \pm 0.0010$ & 8.21 & 9.6 \\
\hline
0.5 & $0.0101 \pm 0.0009$ & 8.20 & 11.7 \\
\hline
0.75 & $0.0102 \pm 0.0009$ & 8.21 & 13.8 \\
\hline
1 & $0.0101 \pm 0.0010$ & 8.20 & 18.6 \\
\hline
1.25 & $0.0099 \pm 0.0010$ & 8.20 & 27.2 \\
\hline
1.5 & $0.0098 \pm 0.0009$ & 8.19 & 46.1\\
 \hline \hline
\end{tabular}
\end{table}


\section{Best-fit parameters}
\label{best_fit}

Here for completeness I provide the posterior probability distributions obtained for the BH-only model and the BH+spike model with an outer halo of slope $\gamma = 0.25$.

\section{Parameters, observables, and analytic solution}

\begin{figure}[t!]
\centering
\includegraphics[width=\linewidth]{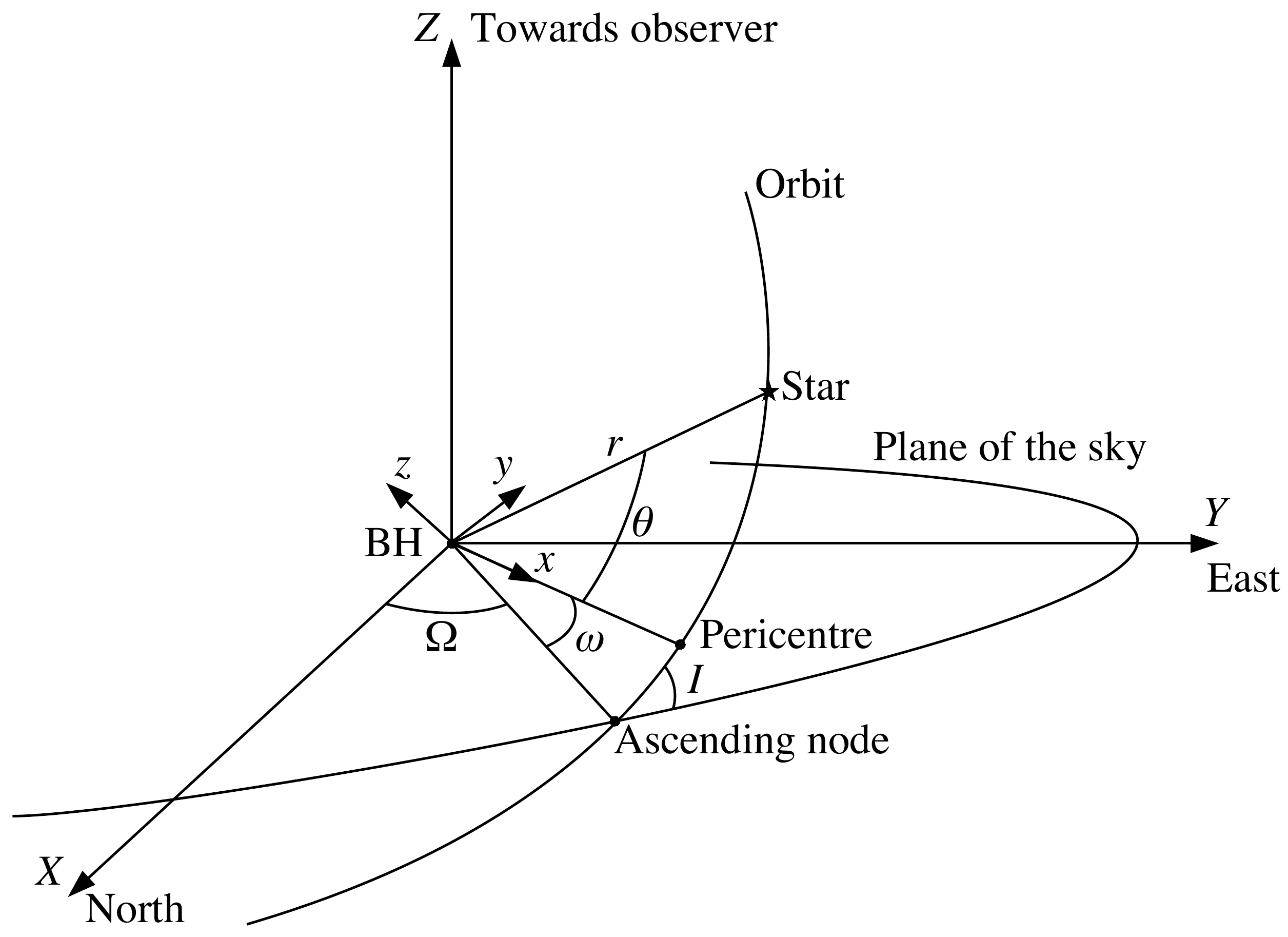} 
\caption{\label{coordinate_system}Definition of the coordinate systems.}
\end{figure}

Here I recall textbook results of standard mechanics in a central potential (e.g., \citealp{Bate1971}). The geometry of the problem is described in Fig.~\ref{coordinate_system}. The orbit of the star lies on a plane due to conservation of angular momentum. Let $r$ and $\theta$ be the polar coordinates of the star on that plane, and $x$, $y$, $z$ the associated Cartesian coordinates. For a stellar orbit we have $x = r \cos \theta$, $y = r \sin \theta$ and $z = 0$. The $x$ axis is defined in such a way that it points towards the pericentre of the orbit.\footnote{We note that elliptical orbits are only relevant in the Keplerian BH-only case.} The dynamics in the orbital plane is well known from first principles. However, observations of the position and velocity of the star are made on the plane of the sky orthogonal to the line of sight pointing to the central object of interest, namely the SMBH Sgr A*. Therefore, we need to transform the phase-space coordinates of the star to a reference coordinate system associated with the plane of the sky. Let $X$, $Y$ and $Z$ be the Cartesian coordinates in that reference system. Conventionally, the positive $X$ direction points to the North and the positive $Y$ direction to the East, while the positive $Z$ direction points to the Earth. 
As shown in Fig.~\ref{coordinate_system}, the orbital plane is inclined with respect to the plane of the sky at an angle $I$. The orbit of the star crosses the plane of the sky in the positive $Z$ direction at the so-called ascending node. The angle between the directions of the ascending node and the pericentre is denoted as $\omega$. Finally, the angle between the $X$ axis---which serves as a reference direction---and the direction of the ascending node is referred to as the longitude of the ascending node, $\Omega$.

As shown in Fig.~\ref{coordinate_system}, the coordinates associated with the plane of the sky are connected to the coordinates in the plane of the orbit via three rotations characterized by $\omega$, $I$ and $\Omega$: first a rotation through $-\omega$ around the $z$ axis, then a rotation through $I$ around the axis connecting the central object and the ascending node, and a rotation through $-\Omega$ around the $Z$ axis. This is expressed in matrix form as:
\begin{align}
\begin{pmatrix}
  X \\
  Y \\
  Z
 \end{pmatrix} & = \begin{pmatrix}
  \cos \Omega & -\sin \Omega & 0 \\
  \sin \Omega & \cos \Omega & 0 \\
  0 & 0 & 1
 \end{pmatrix} 
  \begin{pmatrix}
  1 & 0 & 0 \\
  0 & \cos I & \sin I \\
  0 & -\sin I & \cos I
 \end{pmatrix} \nonumber \\
 & \times \begin{pmatrix}
  \cos \omega & -\sin \omega & 0 \\
  \sin \omega & \cos \omega & 0 \\
  0 & 0 & 1
 \end{pmatrix}
\begin{pmatrix}
  x \\
  y \\
  z
 \end{pmatrix}.
\end{align}
This can be rewritten as the following functions of time $t$:
\begin{align}
\label{X}
X(t) & = r(t) \left[ \cos \Omega \cos \left( \omega + \theta (t) \right) - \sin \Omega \sin \left( \omega + \theta (t) \right) \cos I \right] \\
\label{Y}
Y(t) & = r(t) \left[ \sin \Omega \cos \left( \omega + \theta (t) \right) + \cos \Omega \sin \left( \omega + \theta (t) \right) \cos I \right] \\
Z(t) & = -r(t) \sin \left( \omega + \theta (t) \right) \sin I,
\end{align}
where I have used $x(t) = r(t) \cos \theta(t)$, $y(t) = r(t) \sin \theta(t)$ and $z(t) = 0$ for the star. Now, the actual observables are angular coordinates on the sky---more specifically right ascension $\alpha$ and declination $\delta$---as well as radial velocity $v_{\mathrm{r}}$. By definition, right ascension and declination are related to the $X$ and $Y$ coordinates on the sky via $\alpha = Y/R_{0}$ and $\delta = X/R_{0}$, where $R_{0}$ is the distance between Earth and Sgr A*.\footnote{Note that we are in the regime of small angles.} The motion of the BH is then accounted for through a linear term in the angular position of the star as a function of time. The observable angular coordinates of the star as a function of time ($\alpha_{*}(t)$, $\delta_{*}(t)$) are thus given by
\begin{align}
\label{alpha}
\alpha_{*}(t)  & = \alpha_{*/\mathrm{BH}}(t) + \alpha_{\mathrm{BH}} + v_{\alpha,\mathrm{BH}} (t - t_{\mathrm{ref}}),\\
\label{delta}
\delta_{*}(t)  & = \delta_{*/\mathrm{BH}}(t) + \delta_{\mathrm{BH}} + v_{\delta,\mathrm{BH}} (t - t_{\mathrm{ref}}),
\end{align}
where $\alpha_{*/\mathrm{BH}}(t) = Y(t)/R_{0}$ and $\delta_{*/\mathrm{BH}}(t) = X(t)/R_{0}$ are the angular coordinates of the star in the frame of the BH---with $X(t)$ and $Y(t)$ given by Eqs.~(\ref{X}) and (\ref{Y}), respectively---and $t_{\mathrm{ref}}$ is a reference time, taken to be 2009 yr as in \citet{Gillessen2017}. The motion of the BH also causes a shift in radial velocity of the star which reads
\begin{equation}
\label{radial_velocity}
v_{\mathrm{r},*,\mathrm{IR}}  = - \dot{Z}_{*} + v_{\mathrm{r,BH}},
\end{equation}
where $\dot{} \equiv \mathrm{d}/\mathrm{d}t$.\footnote{Note the minus sign in Eq.~(\ref{radial_velocity}) which accounts for the definition of the $Z$ axis.} \footnote{Measured radial velocities are given in the reference frame of the LSR.} 

\begin{figure*}[t!]
\centering
\includegraphics[width=\linewidth]{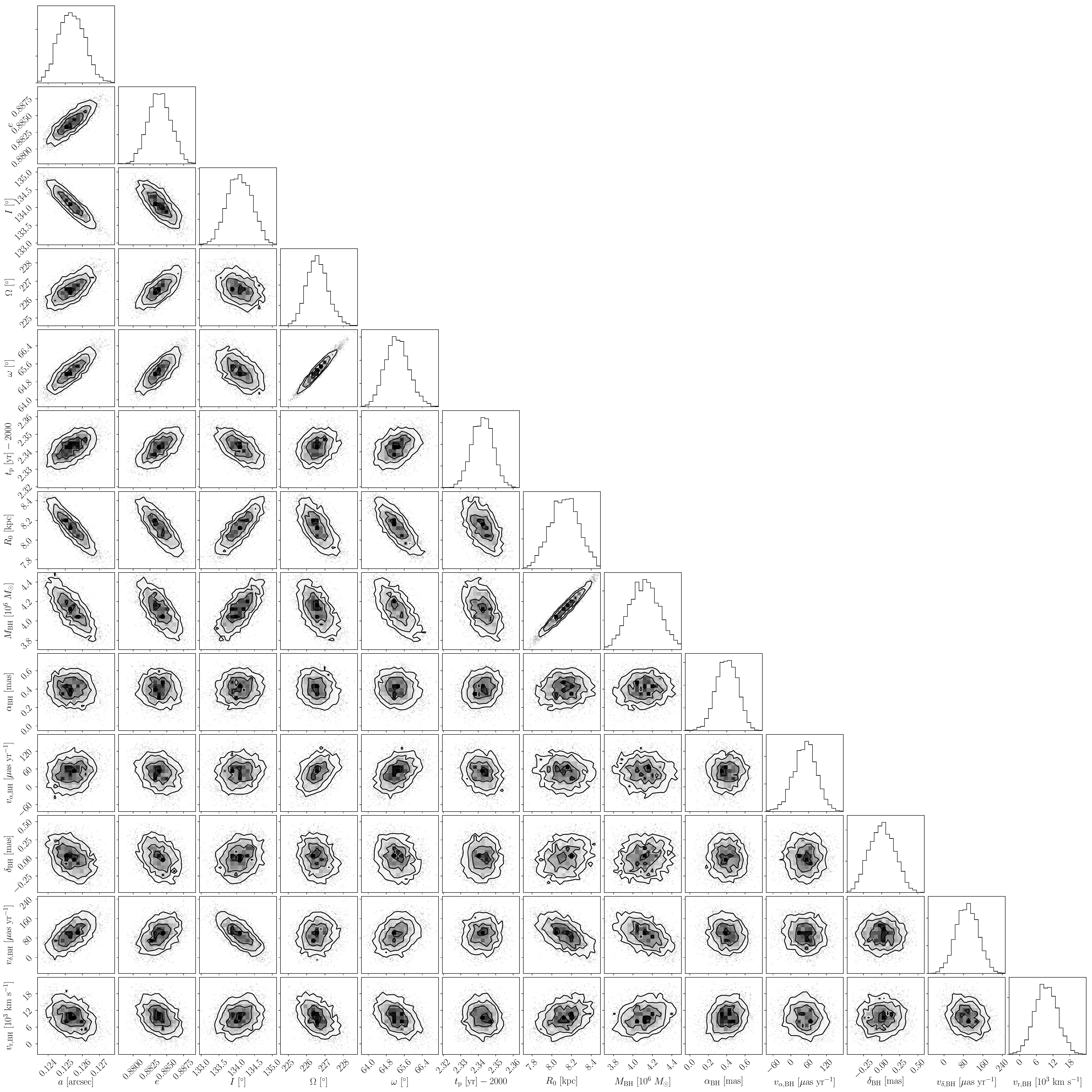}
\caption{\label{corner_plot_BH_only_VLT_2016}Marginalized posterior probability density functions for the 13 parameters of the BH-only model, using the entire VLT data set up to 2016. This scatterplot matrix was produced using the \textsf{corner.py} Python module \citep{Foreman-Mackey2016}.}
\end{figure*}

\begin{figure*}[t!]
\centering
\includegraphics[width=\linewidth]{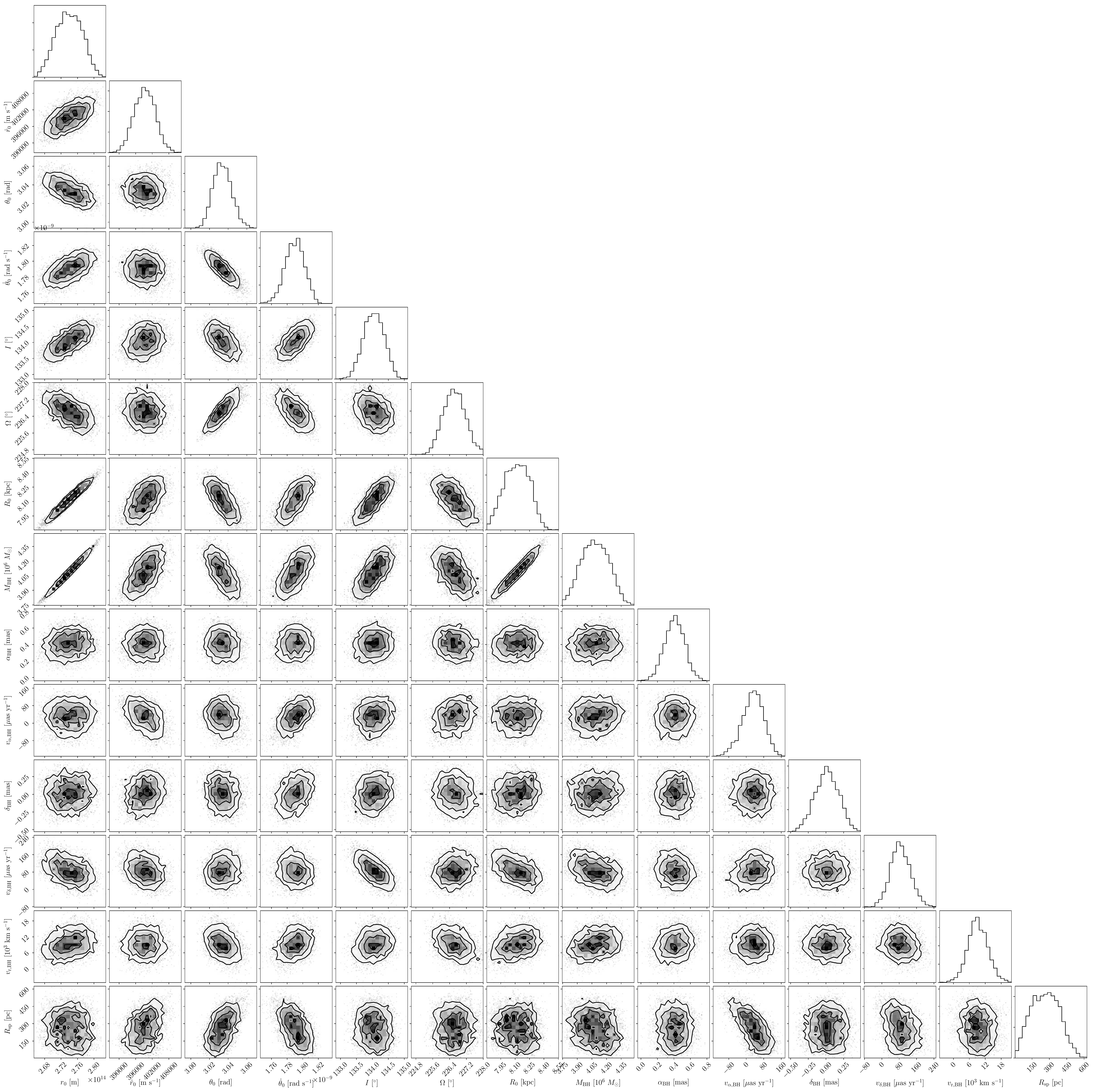}
\caption{\label{corner_plot_BH+spike_VLT_2016}Marginalized posterior probability density functions for the 14 parameters of the BH+spike model (for fixed halo slope $\gamma = 0.25$), using the entire VLT data set up to 2016. This scatterplot matrix was produced using the \textsf{corner.py} Python module \citep{Foreman-Mackey2016}.}
\end{figure*}

Now I recall the main equations that characterize the solution of the equations of motion of the star when the gravitational potential is only created by the central SMBH. The time dependence of $r$ and $\theta$ (or equivalently $x$ and $y$) in the plane of the orbit is expressed through the elliptic anomaly $E(t)$ defined by
\begin{equation}
\label{r(t)}
r(t) = a (1 - e \cos E(t)),
\end{equation}
and solution of the implicit Kepler equation
\begin{equation}
\label{E(t)}
E(t) - e \sin E(t) = \mathcal{M}(t),
\end{equation}
where the so-called mean anomaly is $\mathcal{M}(t) = n (t-t_{\mathrm{p}})$, with $n=2\pi/T$ and the orbital period $T$ of the star is obtained from Kelper's third law:
\begin{equation}
T = 2\pi \left( \dfrac{a^{3}}{G M_{\mathrm{BH}}} \right) ^{1/2}.
\end{equation}
The time dependence of $\theta$ is given by 
\begin{equation}
\label{theta(t)}
\theta(t) = 2 \arctan \left[ \left( \dfrac{1+e}{1-e} \right)^{1/2} \tan \left( \dfrac{E(t)}{2} \right)  \right],
\end{equation}
Once $E(t)$ is known, the evolution of the orbit is entirely characterized by Eqs.~(\ref{r(t)}) and (\ref{theta(t)}). Eq.~(\ref{E(t)}) does not have a simple analytic solution but can be solved iteratively. The derivatives of the $r$ and $\theta$ are also useful:
\begin{equation}
\label{r_dot}
\dot{r}(t) = \dfrac{n a^{2} e}{r(t)} \sin E(t)
\end{equation}
and
\begin{equation}
\label{theta_dot}
\dot{\theta}(t) = \dfrac{\left( 1-e^{2} \right)^{1/2} n a^{2} }{r(t)^{2}}.
\end{equation}
The numerical resolution is then compared with the analytic solution of Eqs.~(\ref{r(t)}) and (\ref{theta(t)}) to calibrate the numerical method that I subsequently apply to the more general problem of an extended mass. For the numerical solution I use Eqs.~(\ref{r(t)}), (\ref{theta(t)}), (\ref{r_dot}) and (\ref{theta_dot}) to compute the initial guesses for the free parameters $r_{0}$, $\theta_{0}$, $\dot{r}_{0}$ and $\dot{\theta}_{0}$ in the minimization procedure.

For the numerical model---crucial when accounting for a DM component---the observables are written in the following way:
\begin{align}
\alpha_{*}(t) & = \alpha_{*/\mathrm{BH}}(t) + \alpha_{\mathrm{BH}},\\
\delta_{*}(t) & = \delta_{*/\mathrm{BH}}(t) + \delta_{\mathrm{BH}},\\
v_{\mathrm{r},*}(t) & = [\dot{r}(t) \sin \left( \omega + \theta (t) \right) + r(t) \dot{\theta}(t) \cos \left( \omega + \theta (t) \right)] \sin I,
\end{align}
with $\dot{\theta}(t) = L/r(t)^{2}$.

\end{document}